\documentclass[prl,twocolumn,showpacs,preprintnumbers,amsmath,amssymb]{revtex4}

\usepackage{graphicx}
\graphicspath{{DiffMass/}}
\usepackage{dcolumn}
\usepackage{bm}
\newcommand{\sgn}{\textrm{sgn}}

\begin{document}

\title{Scattering and binding of different atomic species in a
  one-dimensional optical lattice} 

\author{Rune T. Piil, Nicolai Nygaard, and Klaus M\o lmer}

\affiliation{Lundbeck Foundation Theoretical Center for Quantum System
  Research, Department of Physics and Astronomy, University of Aarhus,
  DK-8000 {\AA}rhus C, Denmark}

\date{\today}

\begin{abstract}
  The theory of scattering of atom pairs in a periodic potential is
  presented for the case of different atoms.  When the scattering
  dynamics is restricted to the lowest Bloch band of the periodic
  potential, a separation in relative and average discrete coordinates
  applies and makes the problem analytically tractable. We present a
  number of results and features, which differ from the case of
  identical atoms.
\end{abstract}


\maketitle

The combination of external confinement and tuning of atomic levels
and molecular potential curves by external fields has led to a rich
variety of experiments with cold atoms, aiming at the study of
formation of ultracold molecules, mean field dynamics, generation of
complex many-body states, and production of quantum correlated sources
of atoms.  While initial studies dealt with only a single atomic
species, the achievement of sympathetic cooling, mixing and
observation of spatial separation of bosonic and fermionic mixtures,
and formation of heteronuclear molecules with permanent dipole moments
have spurred activities on mixtures of atomic species
\cite{Deuretzbacher08,Nemitz08,Inouye04,Stan04,Wille08}.

In this paper, we consider the physics of a pair of atoms labeled A
and B in an optical lattice consisting of three standing wave laser
beams intersecting at right angles with wavelength $\lambda_L$. The
atoms experience light induced energy shifts, which result in a cubic
periodic potential $V_{\mathrm{lat}, \beta}(\mathbf{x})=
\sum_{i=1,2,3}V_{\beta }^{i}\sin^2(\pi x_i/a)$, where $a=\lambda_L/2$
is the lattice constant and $V_{\beta }^i$ is the lattice strength for 
the atomic species $\beta=A,B$ along direction $i$. The lattice
strengths depend on the polarizability of the atoms as well as the
amplitude and detuning of the electromagnetic field.  In our numerical
examples we apply the physical parameters relevant to $^{40}$K and
$^{87}$Rb atoms, studied in Ref. \cite{Deuretzbacher08}. We take A to
be $^{87}$Rb and B to be $^{40}$K with
$V^3_\mathrm{A}=3E_\mathrm{R,Rb}$,
$V^1_\mathrm{A}=V^2_\mathrm{A}=40E_\mathrm{R,Rb}$ and
$V^i_\mathrm{B}=0.86V^i_\mathrm{A}$, where
$E_{\rm{R},\mathrm{Rb}}={h^2}/{2m_\mathrm{Rb}\lambda_{\rm{L}}^2}$ is
the recoil energy for rubidium.

The motion of a single atom is governed by the Hamiltonian
$\mathcal{H}_0^\beta=-(\hbar^2/2m_\beta)\nabla^2 + V_{\mathrm{lat},
  \beta}(\mathbf{x})$, which separates into three independent
one-dimensional equations, each solved by Bloch wave functions
$\phi_{nq}^\beta(x)$, where $n$ is the band index and $q\in[-\pi/a,
\pi/a]$ is the quasi-momentum. Within each band the Bloch waves can be
transformed into the localized Wannier basis functions
$w_{nz_j}^\beta(x)$ centered around the potential minima $z_j=ja$
\cite{Kohn59}. We are interested in the quasi-one-dimensional regime,
which is obtained when the lattice potential in one
(longitudinal) direction is much weaker than in the two other
(transverse) directions $V_{\beta}^{3}\ll
V_{\beta}^{1}=V_{\beta}^{2}$. In this situation the transverse motion
will be confined to the ground state of an effective harmonic
oscillator with frequency
$\omega_\beta=({\pi}/{a})\sqrt{2V_\beta^{1}/m_\beta}$ and eigenstate
of motion $w^\beta(x)=\left( {m_\beta\omega_\beta}/{\pi\hbar}
\right)^{1/4} \exp(-{m_\beta\omega_\beta x^2}/{2\hbar})$ whenever the
longitudinal motional energies are significantly below
$\hbar\omega_\beta$. The  motion is then effectively one-dimensional along the direction $x_3$.

The dynamics of a single atom in a one-dimensional lattice is
described by the dynamical tunneling amplitudes $J_\beta = \langle
w_{nz_j}^\beta|\mathcal{H}_0^\beta|w_{nz_{j+1}}^\beta \rangle$. In the
following we only include nearest-neighbor tunneling, and we consider
motion restricted to the lowest Bloch band ($n=1$), but our analysis
and results may be generalized to higher bands and beyond
nearest-neighbor tunneling~\cite{Piil07}.


We characterize the two-body system by a wavefunction, which is
expanded on Wannier product wavefunctions
$\Psi(x_{3,A},x_{3,B})=\sum_{z_A,z_B}\psi(z_A,z_B) w_{nz_A}^A(x_{3,A})
w_{nz_B}^B(x_{3,B})$ and solve for the amplitude $\psi(z_A,z_B)$ of
finding atom A and atom B in the Wannier functions centered at the
discrete sites $z_A$ and $z_B$, respectively.  In this basis the
two-body non-interacting Hamiltonian becomes
\begin{equation}
  \label{eq:H0}
  {H}_0=-J_A(\Delta_{z_A}+2)-J_B(\Delta_{z_B}+2),
\end{equation}
where $\Delta_{z}f(z) = f(z+a)+f(z-a)-2f(z)$ is the discrete
Laplacian.  The eigenstates of $H_0$ are product states
\begin{equation}
  \label{eq:psi_0}
  \psi(z_A,z_B)=\exp(iq_Az_A)\exp(iq_Bz_B)
\end{equation}
with energy 
\begin{equation}
  \label{eq:E0_I}
  \epsilon(q_A,q_B)=E_A(q_A) + E_B(q_B),
\end{equation}
where $E_\beta(q_\beta)=-2J_\beta\cos(q_\beta a)$ is the single
particle energy dispersion.  For convenience we put the zero of energy
in the middle of the first Bloch band.

The eigenstate in Eq. (\ref{eq:psi_0}) is a product of Bloch waves
with quasi-momenta $q_A$ and $q_B$. We now change to collective
$Z=(z_A+z_B)/2$ and relative $z=z_A-z_B$ coordinates, and we introduce
the collective $K=q_A+q_B$ and relative $q=(q_A-q_B)/2$ quasi-momenta.
In the special case of equal masses the collective coordinate is
identical to the center-of-mass coordinate. Due to the discrete nature
of the problem the Hamiltonian separates into a collective and a
relative coordinate part via the Ansatz
$\psi(z_A,z_B)=e^{iKZ}\psi_K(z)$. By applying ${H}_0$ to the product
Ansatz we get $H_0e^{iKZ}\psi_K(z)=e^{iKZ}{H}^0_K\psi_K(z)$, where
$H_K^0$ is the action of the Hamiltonian $H_0$ on the relative
coordinate part $\psi_K$ given by
\begin{equation}
  \label{eq:H0Z}
  \begin{split}
    {H}^0_K\psi_K(z)=-J_A\left(e^{iKa/2}\psi_K(z+a)+
      e^{-iKa/2}\psi_K(z-a)\right)\\
    -J_B\left(e^{-iKa/2}\psi_K(z+a)+e^{iKa/2}\psi_K(z-a)\right).
  \end{split}
\end{equation}
Note that the relative motion Hamiltonian ${H}^0_K$ acts on components
of the joint system, where $K$ has a specified value. 

Unlike the case of identical atoms, where $J_A=J_B$, ${H}^0_K$ is not
invariant under complex conjugation, and hence is not time-reversal
invariant, for $K\ne 0$. The collective quasi-momentum is here playing
a role similar to that of a classical magnetic field on the motion of
electrons in an atom or a solid. As a consequence of the breaking of
the time-reversal symmetry, we cannot expect the bound state
wavefunctions to be real, but in further analogy with the magnetic
interactions we note that time reversal of the full two-body dynamics
is accompanied by a change of sign of $K$ (of the magnetic field), and
hence ${H}^0_K=({H}^0_{-K})^*$. If $\psi_{-K}$ is an eigenstate of
${H}^0_{-K}$ then the complex conjugate, and hence time-reversed, wave
function $\psi_{-K}^{*}$ is an eigenfunction of ${H}^0_K$ with the
same eigenvalue.

\begin{figure}
  \includegraphics[width=\columnwidth]{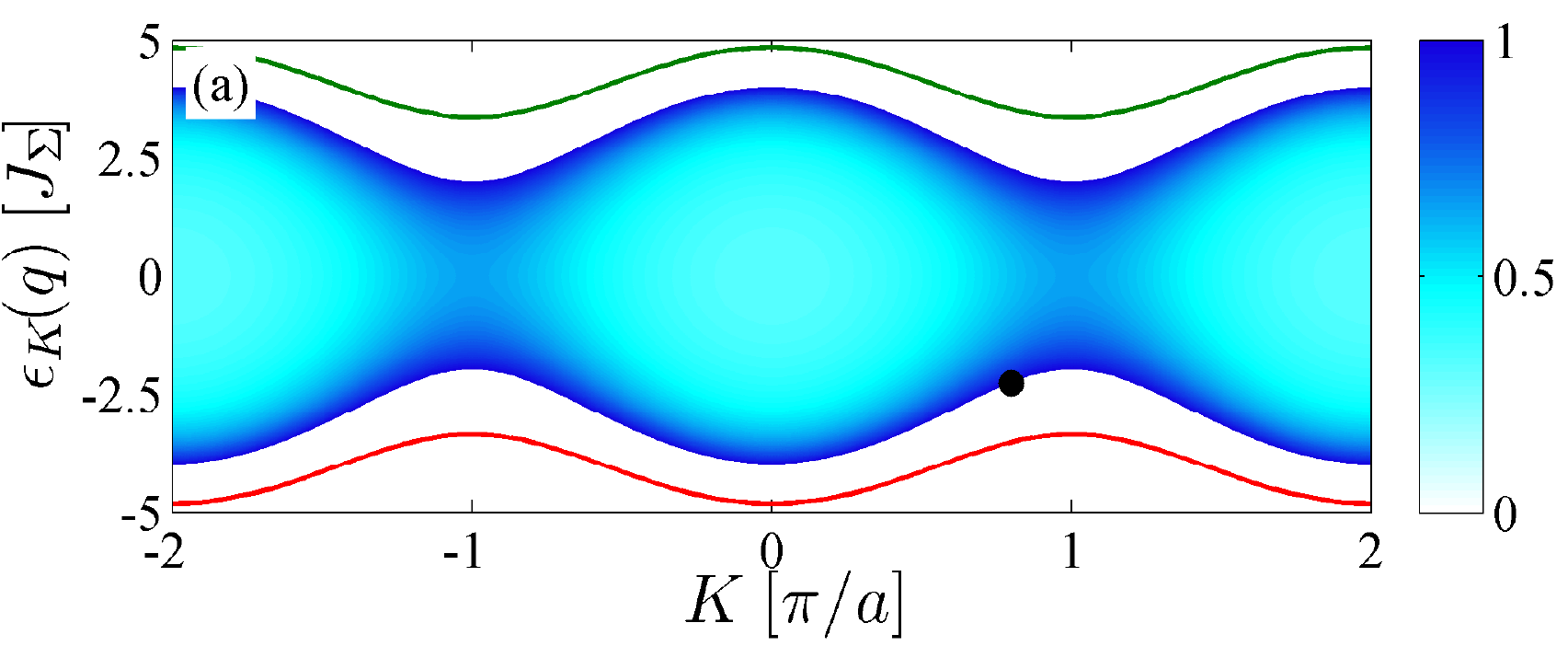}
  \includegraphics[width=\columnwidth]{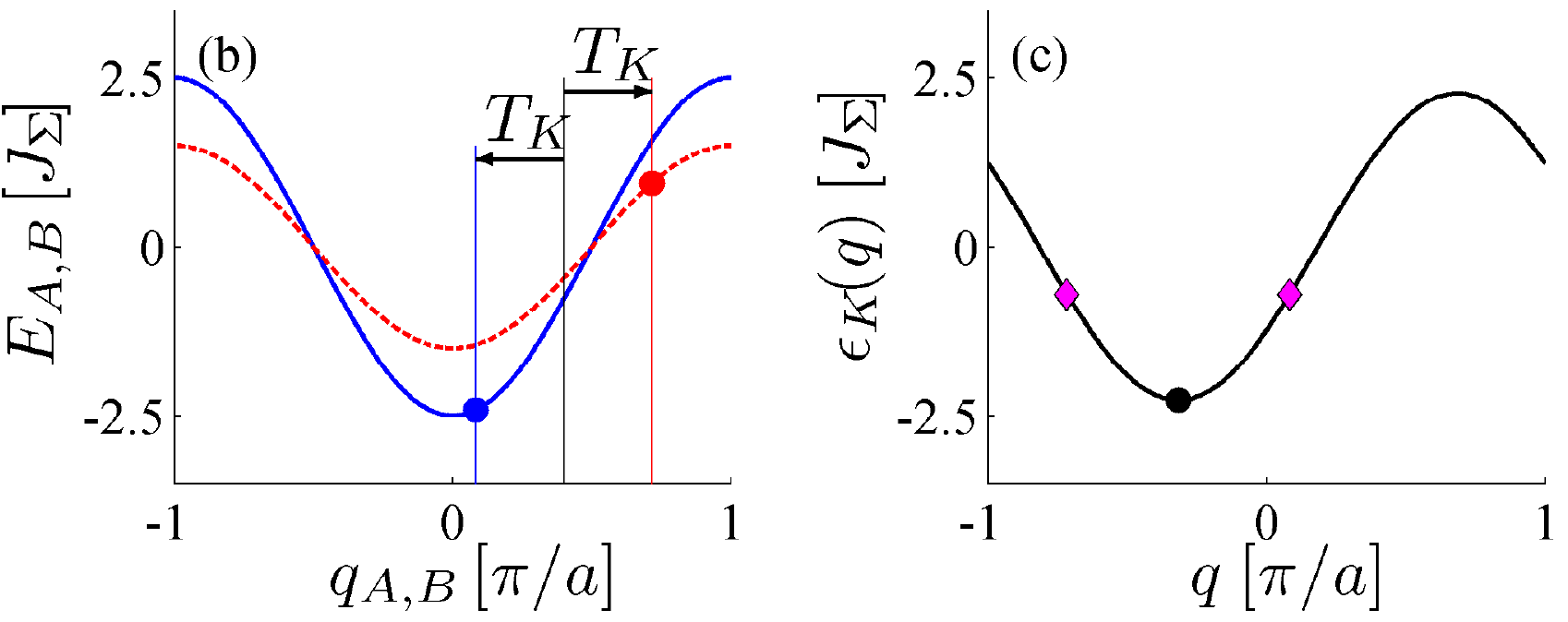}
  \caption{(a) Energy band spanned by $\epsilon_K(q)$ for the ratio
    $J_\Delta=0.5J_\Sigma$. The single atom spectra $E_\beta(q_\beta)$
    in (b) and the two-body spectrum
    $\epsilon_K(q)=E_A(K/2+q)+E_B(K/2-q)$ in (c) are shown for
    $K=0.8\pi/a$. The corresponding energy minimum
    $(K,q)=(0.8\pi/a,-T_K)$ is marked by dots in all three figures and
    the vertical lines in (b) indicate from left to right $q_A$, $K/2$
    and $q_B$. The lower and upper curve in (a) shows the bound state
    energy for $U=-2.7J_\Sigma$ and $U=2.7J_\Sigma$, respectively,
    while the shading shows the reflection coefficient for
    $|U|=2.7J_\Sigma$ [see Eq.~(\ref{eq:Rbg})]. The significance of
    the diamonds in (c) is explained after Eq.~(\ref{eq:psi_bg}).}
  \label{fig:EK}
\end{figure}

Introducing the average and half-difference tunneling amplitudes,
$J_{\Sigma,\Delta}=(J_A \pm J_B)/2$, with values $J_\Sigma=J_A=J_B$
and $J_\Delta=0$ in the case of identical particles, the energies
(\ref{eq:E0_I}) can be written in the convenient form
\begin{equation}
  \label{eq:E0}
  \epsilon_K(q)=E_K\cos\left[(T_K+q)a\right],
\end{equation}
which is parametrized by the collective energy
\begin{equation}
  \label{eq:E_K}
  E_K=-4\sqrt{J_\Sigma^2\cos^2(Ka/2)+J_\Delta^2\sin^2(Ka/2)}
\end{equation}
and by the quasi-momentum shift
\begin{equation}
  \label{eq:TK}
  T_Ka=\sin^{-1}\frac{4J_\Delta\sin(Ka/2)}{|E_K|},
\end{equation}
where $T_Ka$ is taken to be in the interval $(-\pi/2,\pi/2]$ for
$Ka\in(-\pi,\pi]$. The collective energy $\pm|E_K|$ determines the
extremal values of the band of energies obtained by varying $q$, and
therefore the total width of the band is $2|E_K|$. The continuum band
is shown as a the colored region in Fig. \ref{fig:EK}(a). For the Rb-K
system and our choice of optical lattice parameters the tunneling
amplitudes have the values $J_\Sigma=0.5$kHz and
$J_\Delta=0.5J_\Sigma$.


There are two significant differences from the identical particle
case: (\textit{i}) the width of the continuum band does not vanish for
$K=\pi/a$, but maintains a finite width of $8|J_\Delta|$. For atom
pairs with $K=0$ the width of the band is $8|J_\Sigma|$. (\textit{ii})
The minimum and maximum of the continuum are obtained for relative
quasi-momenta $q=-T_K$ and $q=\pm\pi/a-T_K$, respectively, and not, as
for identical particles, at relative quasi-momenta $0$ and $\pm\pi/a$.

The finite quasi-momentum difference at the band edges,
$q_A-q_B=-2T_K\pmod{2\pi/a}$ , appears because the energy dispersions
of the atoms have different amplitudes, $2J_\beta$. This point is
illustrated in Fig. \ref{fig:EK}(a)-(c), where the lowest energy state within the band  
for $Ka/\pi=0.8$ is illustrated by dots. The quasi-momenta of
the individual atoms are displaced by $\pm T_K$ from $K/2$.

\begin{figure}
  \includegraphics[width=\columnwidth]{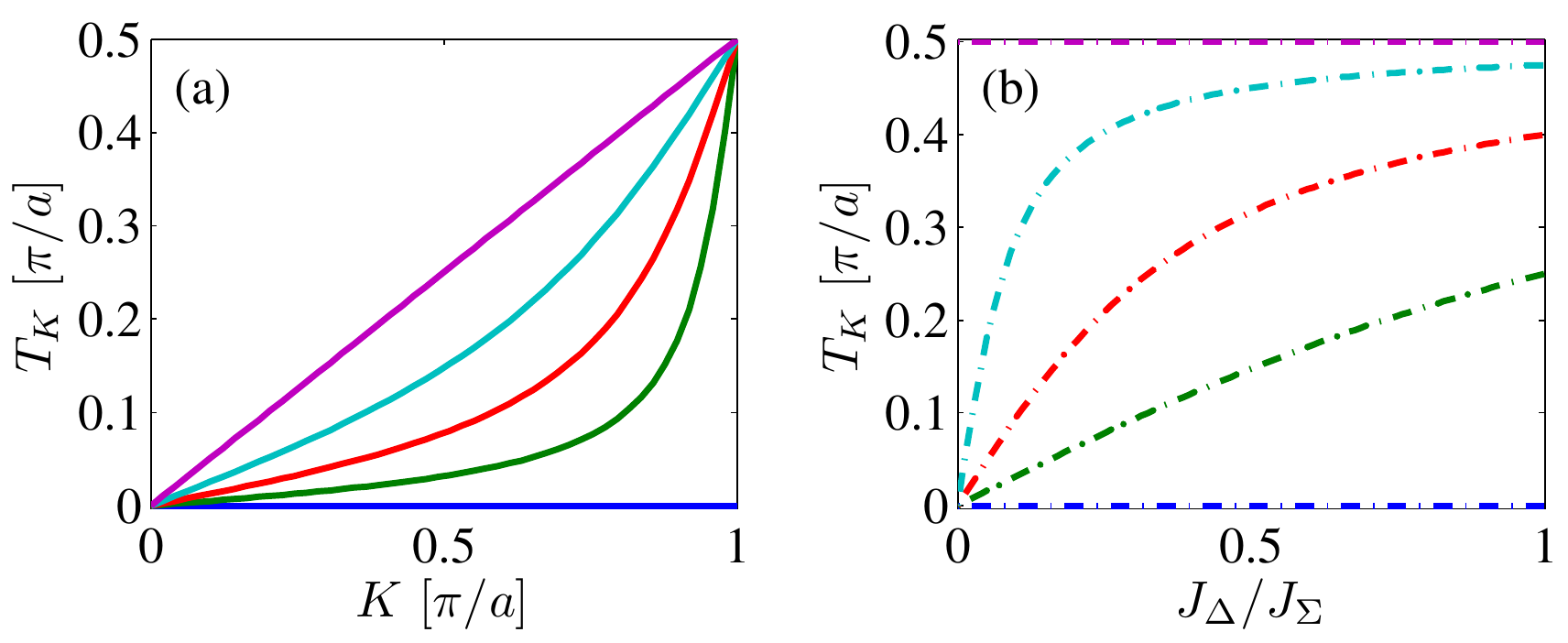}
  \caption{Shift of the relative momentum, $T_K$. (a) From below
    $J_\Delta/J_\Sigma=0,\ 0.1,\ 0.25,\ 0.5,\ 1$. (b) From below
    $Ka/\pi=0,\ 0.5,\ 0.8,\ 0.95,\ 1$.}
  \label{fig:TK}
\end{figure}

The value of $T_K$ is depicted in Fig. \ref{fig:TK}. $T_K$ is an odd
function of both $K$ and $J_\Delta$. In Fig. \ref{fig:TK}(a) the lower
curve $J_\Delta/J_\Sigma=0$ corresponds to $J_A=J_B$, the identical
particle case, where $T_K=0$ for all $K$ because both atoms have the
same energy dispersion, and the upper curve $J_\Delta/J_\Sigma=1$
corresponds to $J_B=0$, where only atom $A$ is allowed to move, and
therefore $q_A=K$, $q_B=0$ and $T_K=K/2$. From Fig. \ref{fig:TK}(b) we
note that $T_{0}$ is always zero, and whenever $J_A\ne J_B$ we have
$T_{\pi/a}=\pi/2a$ due to the symmetry of the single particle
dispersions. 

So far, our analysis has only accounted for the separable state
(\ref{eq:psi_0}) of non-interacting atoms in a collective set of
coordinates~\cite{Note1}. In the presence of an atomic interaction
depending only on the relative coordinate the separation in $z$ and
$Z$ is maintained, and we now turn to the description of the dynamics
and bound states of the interacting system. In the Wannier basis we
assume the interaction potential, $\hat{U}$, to be on-site,
\textit{i.e.}, of the form $U\delta_{z,0}$, where $U$ is a function of
the three-dimensional background scattering length and the optical
lattice parameters.  The bound and scattering states are analyzed
using the relative motion Green's function for the non-interacting
particles, $\hat{G}_K^0(E)=[E-\hat{H}_0]^{-1}$. Due to the assumption
of on-site interactions, we only need to specify $G_K^0(E,z)=\langle
z|\hat{G}_K^0(E)|0\rangle$ given by the Fourier transform
\begin{equation}
  \label{eq:G0}
  G_K^0(E,z)=\int^{\pi/a}_{-\pi/a} \frac{dq}{2\pi}
  \frac{ae^{iqz}}{E-E_K\cos((T_K+q)a)+i\eta}
\end{equation}
of the relative quasi-momentum Green's function,
$\mathcal{G}_K(E;q,q')=\delta(q-q')/(E-\epsilon_K(q))$.  Here $\eta$
is a positive infinitesimal added to enforce outgoing boundary conditions. 
By a change of coordinate $q\to q-T_K$ the
integral is identical, except for a front factor $\exp(-iT_Kz)$, to
the one studied previously for identical particles
\cite{Nygaard08b}. Hence, for energies inside the continuum band we
have
\begin{equation}
  \label{eq:G0in}
  G_K^0(E,z)=-\frac{ie^{-iT_Kz}e^{ip|z|}}{\sqrt{E_K^2-E^2}},
\end{equation}
where $pa=\cos^{-1}(E/E_K)$, and outside the continuum we find
\begin{equation}
  \label{eq:G0out}
  G_K^0(E,z)=\sgn(E)
  \frac{e^{-iT_Kz}e^{-\kappa|z|}}{\sqrt{E^2-E_K^2}} [-\sgn(E)]^{z/a}
\end{equation}
with $\kappa a=\cosh^{-1}|E/E_K|$. The Green's function is similar to
the identical particle case, except for a complex factor $e^{-iT_Kz}$,
as already mentioned, and a modified expression for $E_K$
(\ref{eq:E_K}).

We use the Dyson equation,
$\hat{G}_K^U(E)=\hat{G}_K^0(E)+\hat{G}_K^0(E)\hat{U}\hat{G}_K^U(E)$,
to find the interacting Green's function
\begin{equation}
  \label{GU_z}
  G^U_K(E,z) = \frac{G^0_K(E,z)}{1-UG^0_K(E,0)},
\end{equation}
and the scattering wavefunction then follows from the
Lippmann-Schwinger equation
\begin{equation}
  \label{eq:psi_bg}
  \psi_K(E,z)=e^{i(p-T_K)z}+UG_K^U(E,0)e^{-iT_Kz}e^{ip|z|}.
\end{equation}
Note that the relative quasi-momenta of the incoming $q=p-T_K$ and
reflected $q'=-p-T_K$ waves are not related in the usual way,
since $q'\ne -q$, unless $K=0$. This is explained by the
degeneracy of $q$ and $q'$ due to the symmetry of the $\epsilon_K(q)$
around the minimum energy state $q=-T_K$, as indicated by the diamonds
in Fig. \ref{fig:EK}(c).  

From the scattered wave in Eq. (\ref{eq:psi_bg}) we can identify the
scattering amplitude $f(E,K)=UG_K^U(E,0)$ and thereby the
reflection coefficient
\begin{equation}
  \label{eq:Rbg}
  R(E,K)=|f(E,K)|^2=\frac{U^2}{E_K^2-E^2+U^2},
\end{equation}
which is indicated by the shading in Fig. \ref{fig:EK}(a). It reaches
unity at the continuum boundaries, where the density of state diverges,
and its minimum at the center of the energy band. The
transmission coefficient $T(E,K)=1-R(E,K)$ might be probed in
scattering experiments or measured spectroscopically by radiative
coupling of a bound molecular state to the continuum
\cite{Nygaard08b}. In the latter case a deep-lying, tightly confined
molecular state $|i\rangle$ is coupled to the continuum by some
transition operator $\hat T$, and the transition probability is then given by
\begin{equation}
  \label{eq:T}
  |\langle \psi_K(E) |\hat{T}| i \rangle|^2\propto
  |\psi_K(E,z=0)|^2=T(E,K). 
\end{equation}

The system supports bound states of the atoms, found as the poles of
the scattering amplitude, $f(E,K)$, with energy
$E_b^0=\sgn(U)\sqrt{E_K^2+U^2}$. As for identical particles, we find a
bound state below the continuum in the case of attractive interactions
and a repulsively bound state \cite{winkler2006rba} above the
continuum in the case of $U>0$. The corresponding bound state
wavefunction of relative motion is given by $G_K^0(E_b^0,z)$,
Eq. (\ref{eq:G0out}), up to a normalization factor. This function is
exponentially decaying with $|z|$.

The bound state wavefunctions are complex with a spatial phase
variation of the relative motion, {\textit{c.f}}
Eq.~(\ref{eq:G0out}). This is in agreement with our earlier discussion
of the lack of time-reversal invariance of the relative motion
Hamiltonian restricted to fixed values of $K$. While such a complex
phase variation may give the impression that one particle is passing
by the other one over and over again within the exponential envelope
of their relative motional state, the two atoms are actually moving
with the same group velocity. To see this, we recall that the bound
states above (below) the continuum have predominantly the
quasi-momentum components of the {\textit{noninteracting}} continuum
states close to the upper (lower) band edge. This implies that the
quasi-momentum distribution of the attractively and repulsively bound
states are peaked around $q=-T_K$ and $q=\pm\pi/a-T_K$, respectively,
which can also be recognized directly from Eq. (\ref{eq:G0out}). For
these eigenstates of $H_0$ a simple calculation
\begin{equation}
  \label{eq:v_AB}
  \frac{\partial\epsilon_K(q)}{\partial
    q}=\frac{\mathrm{d}E_A}{\mathrm{d}q_A}
  \frac{\mathrm{d}q_A}{\mathrm{d}q} + \frac{\mathrm{d}E_B}{\mathrm{d}q_B}
  \frac{\mathrm{d}q_B}{\mathrm{d}q} =\hbar(v_A-v_B)
\end{equation}
relates the group velocities $v_\beta$ of species $\beta$ to the
derivative of the energy dispersion.  At the band edges
$\epsilon_K(q)$ reaches its extrema, \textit{i.e.},
${\partial\epsilon_K(q)}{\partial q}=0$, exactly when $q=-T_K$ or
$q=\pm\pi/a-T_K$, and the group velocities agree for these relative
quasi-momentum components. Furthermore, since $\epsilon_K(q)$ is an even
function around $q=-T_K$, the derivative and hence the difference in
group velocity $v_A-v_b$ is odd.  If we return to the scattering state
in Eq. \eqref{eq:psi_bg}, we therefore observe that the difference in
group velocity always changes sign when the wave is reflected.

\begin{figure}
  \includegraphics[width=\columnwidth]{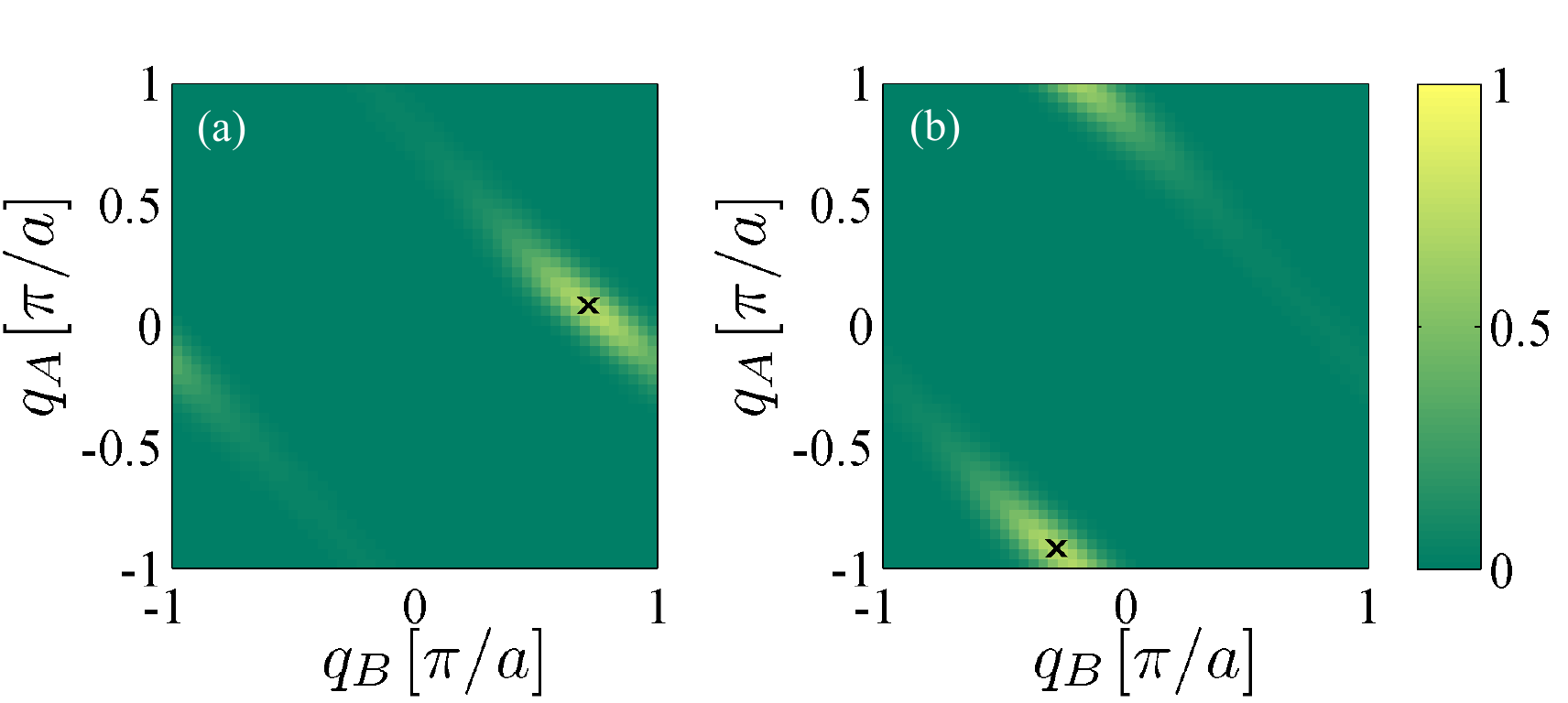}  
  \caption{Quasi-momentum probability distribution in units of
    $(a/2\pi)^2$ for the species $A$ and $B$ in a bound state with
    (a) $U=-2.7J_\Sigma$ and (b) $U=+2.7J_\Sigma$.}
  \label{fig:Quasi-momentum-dist}
\end{figure}

In Fig. \ref{fig:Quasi-momentum-dist} we show the quasi-momentum
distribution of the two components $A$ and $B$ for both a repulsively
(a) and attractively (b) bound atom pair~\cite{Note2}. We have assumed
a Gaussian distribution in $K$ with center $\bar{K}=0.8\pi/a$ and
standard deviation $\sigma_K=0.1\pi/a$. Clear peaks appear at
$q_A+q_B=\bar{K}$ and $q_A-q_B=-2T_{\bar{K}}$ as marked by
crosses. Here $T_{\bar{K}}=0.32\pi/a$, corresponding to
$(q_A,q_B)=(0.08\pi/a,0.72\pi/a)$ in the attractive case (a), while
$(q_A,q_B)=(-0.92\pi/a,-0.28\pi/a)$ in the case of repulsion (b).

\begin{figure}
  \includegraphics[width=\columnwidth]{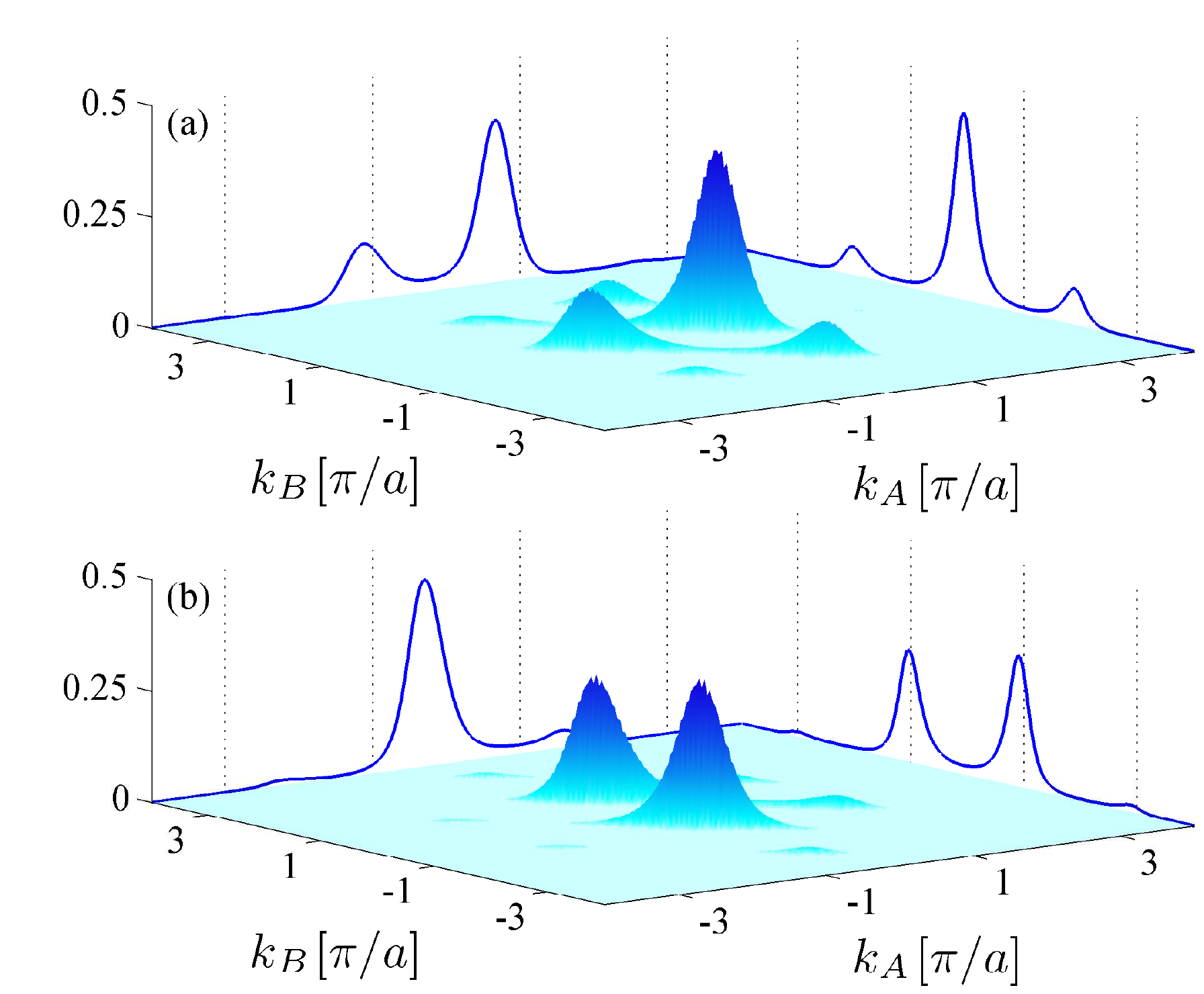}
  \caption{Momentum probability distribution in units of $(a/2\pi)^2$
    for a pair of (a) attractively bound and (b) repulsively bound
    atoms with the same parameters as in
    Fig. \ref{fig:Quasi-momentum-dist}. The solid lines show the
    projected probability of each species $P(k_{A,B})$ in units
    $a/2\pi$.}
  \label{fig:momentum-dist}
\end{figure}

We suggest measuring the wave number shift $T_K$ by rapidly turning
off the lattice potential, because an adiabatic ramp down would alter
$J_\Delta$ and thereby $T_K$. The shift of the main peak in the
momentum spectrum of each individual species is then given by $K/2\to
K/2\pm T_K$ for $U<0$ and $K/2\to K/2\pm T_K\pm \pi/a$ for $U>0$. The
case with attractive interaction is shown in
Fig. \ref{fig:momentum-dist}(a) with the same spread in $K$ as in
Fig. \ref{fig:Quasi-momentum-dist}.  In an experiment the difference
between the main momentum peaks of the two species is $2T_K$. The case
with repulsive interaction is shown in
Fig. \ref{fig:momentum-dist}(b). Alternatively, if the species are
held by different lasers, the lattice may be turned off adiabatically
in a controlled fashion that keeps the difference in the tunneling
amplitudes $J_\Delta$, and hence $T_K$, fixed.

In this paper we have presented an analytical description of both the
dynamics and the bound states of an atom pair of different atoms in a
quasi-one-dimensional lattice.  We suggest several experiments to
probe our model. The transmission coefficient, Eq. (\ref{eq:Rbg}), can
be probed either by a collision experiment or by RF coupling a
strongly bound state to the structured continuum, and the wave number
shift $T_K$ of the bound states can be measured from the momentum
distribution in a time of flight experiment.

We have assumed the two atoms to be in the same Bloch band, but they
could just as well be in different Bloch bands, which also gives rise
to different tunneling amplitudes $J_A\ne J_B$.  In addition, the
model can be extended to describe two-channel magnetic Feshbach
resonances as outlined for identical particles in
Refs. \cite{Nygaard08, Nygaard08b} by replacing $E_K$ and by modifying
the Green's function in those papers in accordance with our
Eqs. (\ref{eq:E_K}), (\ref{eq:G0in}) and (\ref{eq:G0out}).


\end{document}